\documentclass[doublecol]{epl2} 

\usepackage{graphicx}
\usepackage[latin1]{inputenc}
\usepackage{mathrsfs}

\usepackage[colorlinks,bookmarks=true,citecolor=blue,linkcolor=red,urlcolor=blue]{hyperref}
\graphicspath{{./}, {figures/},{../Figs/}}

\title{Symmetry-protected topological phases of alkaline-earth cold fermionic atoms in one dimension}

\author{
   H. Nonne \inst{1} \and
   M. Moliner \inst{2} \and
   S. Capponi \inst{3} \and
   P. Lecheminant \inst{2} \thanks{E-mail: \email{Philippe.Lecheminant@u-cergy.fr}} \and 
   K. Totsuka \inst{4} 
}
\shortauthor{H. Nonne \etal}

\institute{ 
   Department of Physics, Technion, Haifa 32000, Israel. \inst{1} \\
   Laboratoire de Physique Th\'eorique et Mod\'elisation, CNRS UMR 8089,
   Universit\'e de Cergy-Pontoise, Site de Saint-Martin,
   F-95300 Cergy-Pontoise Cedex, France. \inst{2} \\
   Laboratoire de Physique Th\'eorique, CNRS UMR 5152, 
   Universit\'e Paul Sabatier, F-31062 Toulouse, France. \inst{3} \\
   Yukawa Institute for Theoretical Physics, 
   Kyoto University, Kitashirakawa Oiwake-Cho, Kyoto 606-8502, Japan. \inst{4} \\
}


\date{\today}

\abstract{
We investigate the existence of symmetry-protected topological phases
in one-dimensional alkaline-earth cold fermionic atoms with general
half-integer nuclear spin $I$ at half filling.  In this respect, some orbital
degrees of freedom are required. They can be introduced by considering
either the metastable excited state of alkaline-earth atoms
or the $p$-band of the optical lattice. Using complementary
techniques, we show that SU(2) Haldane topological phases 
are stabilised from these orbital degrees of freedom. 
On top of these phases, we find the
emergence of topological phases with enlarged SU($2I+1$) symmetry
which depend only on the nuclear spin degrees of freedom.  The main
physical properties of the latter phases are further studied using a
matrix-product state approach.  
On the one hand, we find that these phases are
symmetry-protected topological phases, with respect to inversion
symmetry, when $I=1/2,5/2,9/2, \ldots$, which is directly relevant to
ytterbium and strontium cold fermions.  
On the other hand, for the other values of $I$(=half-odd integer), these topological phases are 
stabilised only in the presence of exact SU($2I+1$)-symmetry. 
}

\begin{document}

\maketitle

Quantum phases of matter with exotic orderings have attracted 
much interest over the years. Prominent examples are topological phases which do not break any symmetry
and cannot be characterised by local order parameters. Though
all one-dimensional (1D) gapful phases have short-range entanglement,
a stable topological phase can still be defined in 1D by the presence of certain symmetries which
protect the phase~\cite{wengu}. A topological phase of this kind is called 
a symmetry-protected topological (SPT) phase~\cite{wengu,chen}. Otherwise, it can be smoothly 
connected to a trivial gapful phase without any phase transition~\cite{pollman10}.
The Haldane phase~\cite{haldane} of the spin-1 Heisenberg chain
displays striking properties which makes it a paradigmatic example
of featureless gapped topological phases in 1D.  
For a long time, two features have been considered as the defining properties of the `topological' Haldane phase: the existence of hidden non-local string ordering~\cite{dennijs} and the presence of a finite gap above a unique ground state (GS) with periodic boundary conditions while emergent spin-1/2 edge states are liberated at the boundaries of open chains~\cite{hagiwara}.
However, the precise meaning    
as a SPT phase has been recognised quite recently~\cite{wengu} 
and it is now known that the Haldane phase is a relatively robust topological phase protected by 
the presence of at least one of the three discrete symmetries: 
the dihedral group of $\pi$ rotations along the $x,y,z$ axes, time-reversal 
and inversion symmetries~\cite{pollman10,pollman12}.  
Also, there are several proposals to realise the Haldane phase in cold-atom 
systems~\cite{Ripoll-04,Dalla-Torre-06,Nonne_2010, kobayashi.  }

A recent breakthrough has led to the complete classification of SPT 
of 1D spin systems~\cite{classification_1,classification_2,classification_3,classification_4}.
It relies on the determination of the projective representations
of the underlying symmetry group which can be obtained using group cohomology.
In particular, for spin systems with a high SU($N$) symmetry, $N$ distinct topological phases
are expected from this classification~\cite{quella}.
A natural question is the physical realisation of these  SU($N$) 
topological phases starting from realistic fermionic systems.  

In this respect, alkaline-earth like fermionic ultracold atoms seem to be very 
promising since they are the best candidates for experimental realisations of exotic
high-symmetry many-body physics~\cite{Gorshkov2010,cazalilla}. 
In this context, the main interest in those atoms stems from the presence of a 
GS $^1S_0$ (``$g$'') and a metastable excited state
$^3P_0$ (``$e$'') between which transitions are forbidden. 
Moreover, both states have zero electronic angular momentum, so that the nuclear spin $I$ is decoupled from the
electronic spin. The nuclear spin-dependent variation of the scattering lengths
is expected to be smaller than $\sim 10^{-9}$ for the $g$ state and $\sim 10^{-3}$ for the $e$  
state. This results in fermionic systems
with an extended SU($N=2I+1$) symmetry~\cite{Gorshkov2010}.
Such gases of strontium ($^{87}$Sr) atoms ($I=9/2$) or ytterbium ($^{171}$Yb, $^{173}$Yb) 
atoms ($I=1/2,5/2$ respectively)
have been cooled down to reach the quantum degeneracy~\cite{desalvo,fukuhara,taie,taie_2012}.

If one only keeps the $g$ state of the atoms, the effective low-energy 
Hamiltonian of this $N$-component Fermi gas boils down to the SU($N$) Hubbard model \cite{Gorshkov2010}.
However, the gapped Mott-insulating phases of the latter model loaded into a 1D optical 
lattice are  known to be spatially nonuniform for all commensurate fillings~\cite{solyom}. 
In this respect, no 1D topological phases can be formed.

In this Letter, we show, by means of complementary methods, that the situation is radically different 
if one includes orbital degrees of freedom. These degrees of freedom are naturally introduced here by either
considering the metastable excited $e$ state of alkaline-earth atoms~\cite{Gorshkov2010} 
or the two-fold degenerate $p$-band of the 1D optical 
lattice \cite{kobayashi}.  The interplay between orbital and nuclear-spin degrees of freedom gives rise to
different 1D insulating topological phases at half filling ($N$ atoms per site).
Two different classes of topological phases are stabilised when $N$ is even, i.e. $I$ is a half-odd
integer. 
We find Haldane phases with integer orbital pseudospin-$N/2$ and
topological phases with enlarged SU($N$) symmetry that we fully characterise.
In particular, we investigate the topological protection of these phases with respect to
the inversion symmetry by means of a matrix-product state (MPS) approach. 
It is shown that the topological phases of alkaline-earth cold fermions are SPT phases when $N/2$ is odd, i.e.,
$I=1/2,5/2,9/2,\ldots$, which is directly relevant to ytterbium and strontium atoms.

\section{Strong-coupling approach} 
The  effective Hamiltonian  which governs the low-energy properties of the problem is the 
two-orbital SU($N$)  Hubbard model:
\cite{Gorshkov2010,kobayashi,xu}
 \begin{eqnarray}
{\cal H}&=&-t \sum_{i, l \alpha} \left(c_{l \alpha,\,i}^\dag c_{l \alpha,\,i+1}^{}  + H.c.\right)
  -\mu \sum_i n_i 
  + \frac{U}{2} \sum_i n^2_i 
  \nonumber\\
 &+& J \sum_i \left[ (T_i^x)^2 + (T_i^y)^2\right] + J_z \sum_i (T_i^z)^2 ,
  \label{alkaTmodel}
\end{eqnarray}
where $c_{l \alpha,\,i}^\dag$ denotes the fermion creation operator at the $i^{th}$ site
with nuclear-spin states $\alpha=1,\cdots,N$. 
The general orbital index $ l=1,2$ stands either for the $g$ and $e$ states of alkaline-earth like 
fermionic atoms \cite{Gorshkov2010} or for the second double degenerate $p$ orbitals, $p_x$ and $p_y$, of the optical lattice \cite{kobayashi}.
In eq.~(\ref{alkaTmodel}), $n_i = \sum_{l \alpha}  c_{l \alpha,\,i}^\dag c_{l \alpha,\,i}^{}$  denotes
the occupation number on the site $i$, and $2 {\vec T}_i = \sum_{l m \alpha}  c_{l \alpha,\,i}^\dag
{\vec \sigma}_{l m} c_{m \alpha,\,i}^{}$ is the orbital
pseudospin operator (${\vec \sigma}$ being the Pauli matrices). 
For the sake of simplicity, we restrict ourself here to the case where the two orbitals 
play a similar role. As it will be discussed below, 
all topological phases of the problem are robust against small terms which breaks the Z$_2$ orbital symmetry
$1 \leftrightarrow 2$.
On top of the U(1)$_c$ charge symmetry 
($c_{l \alpha,\,i} \rightarrow e^{i \theta} c_{l \alpha,\,i}$), model (\ref{alkaTmodel}) features 
an SU($N$) symmetry ($c_{l \alpha,\,i} \rightarrow \sum_{\beta}  U_{\alpha \beta}  c_{l \beta,\,i}$,
$U$ being an SU($N$)  matrix), and a U(1)$_o$ symmetry in the orbital space spanned 
by the orbital states ($c_{1,2\, \alpha,\,i} \rightarrow e^{\pm i \theta} c_{1,2\, \alpha,\,i}$).
Though model (\ref{alkaTmodel}) enjoys an extended  U(1)$_c$ $\times$  U(1)$_o$ $\times$
SU($N$), the determination of its phase diagram is a highly non-trivial problem in the general case.
However, one can already anticipate the existence of different topological phases by means of  
a strong-coupling analysis along special lines of model (\ref{alkaTmodel}).
These phases are protected by their spectral gaps and are stable against small symmetry-breaking perturbations.
In this respect, let us consider the special case $J=J_z$ where the 
U(1)$_o$ orbital symmetry is enlarged to SU(2)$_o$.  The single-site energy-spectrum at
half-filling ($\mu = U N$) is labelled by two integers $p,q$:
\begin{eqnarray}
E \left(p,q\right) &=& \frac{U}{2} (p+q) \left( p+q - 2 N\right) + \frac{J}{4} \left(p - q\right)(p - q + 2)
\nonumber \\ 
D(p,q) &=& \frac{N! (N+1)! (p-q+1)^2}{(N-p)!(N+1-q)!(p+1)!q!};
   \label{spectrum}
\end{eqnarray}
\begin{figure}
\onefigure[width=0.95\linewidth]{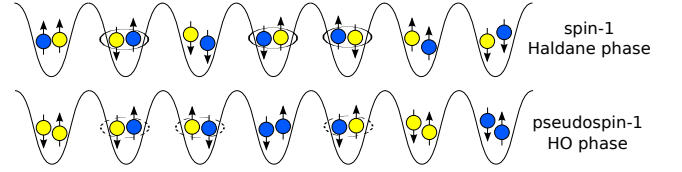}
\caption{(Colour on-line) For $N=2$: the blue and yellow atoms describe the orbital states $l=1,2$. 
The Haldane phases for nuclear spin and orbital (HO) degrees of freedom present a dilute order~\cite{dennijs} 
that can be seen here as an alternation of sites with total (pseudo)spin component $\pm 1$, diluted with an 
arbitrary number of sites with (pseudo)spin component $0$.
\label{fig:HSHO}}
\end{figure}
$D$ is the degeneracy,  $n=p+q$ is the number of fermions on one site,
and $T = (p -q)/2$ is the spin of the  orbital
pseudospin operator ${\vec T}_i$.
A first interesting line is $U=0$ and $J<0$ where the lowest energy states (\ref{spectrum}) are $N+1$
degenerate ($p=N$ and $q=0$) and ${\vec T}_i$ is a $N/2$ pseudospin operator. 
 At second order of perturbation theory in $|J| \gg t$, we find a pseudospin-$N$/2
antiferromagnetic SU(2) Heisenberg chain: ${\cal H}_{\rm eff} = J_o
\sum_i {\vec T}_i \cdot {\vec T}_{i+1}$ with $J_o = 8 t^2/N(2N+1)|J|$. 
As it is well-known, the physics of the latter model strongly depends on the parity 
of $N$~\cite{haldane}. 
When $N$ is odd, the phase is gapless with one gapless bosonic mode while, in the even $N$ case, it 
is fully gapped (Haldane gap).
We thus find the emergence of an SU(2) topological (Haldane) phase when $N$ is even, where the SU(2) symmetry 
stems from the orbital degrees of freedom.  
We dub this phase {\it Haldane-orbital phase} (HO), since this phase is different from 
the Haldane phase for (nuclear) spin degrees of freedom (see fig.~\ref{fig:HSHO}, where a cartoon 
represents this phase for $N=2$).
As it is well known, these Haldane phases are robust against SU(2) symmetry-breaking perturbations. 
In addition, they are SPT phases only when
$N/2$ is odd~\cite{pollman12}.

A second interesting line is $J = 2 NU/(N+2) > 0$ with even $N$.
The lowest energy states correspond to $p=q=N/2$ ($N$ fermions which 
are orbital singlets) and transform into the SU($N$) self-conjugate representations described by a Young tableau with
two columns and $N/2$ rows. At second order of perturbation theory in $|U| \gg t$, 
we find an SU($N$) Heisenberg 
spin chain in this representation. For $N=2$, the spin-1 Haldane phase is 
formed among the nuclear spins  and, in this respect, is intrinsically different from 
the HO phase with $N=2$ (see fig.~\ref{fig:HSHO}).
When $N>2$, the physical properties of the GS of the SU($N$) magnet are not
known. A confinement of spinons is expected from the general classification
of Ref.~\cite{greiter}, and a non-degenerate gapful phase was predicted in the 
large $N$ limit~\cite{sachdev}.

\section{Valence-bond-solid (VBS) approach}
\begin{figure}
\onefigure[width=0.95\linewidth]{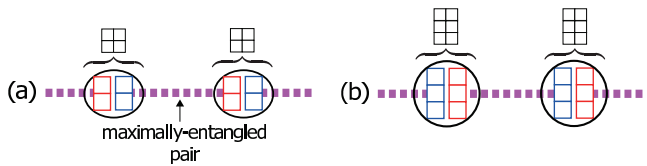}
\caption{(Colour on-line) SU($N$) VBS states are constructed out of a pair of self-conjugate 
representations at each site. Dashed lines denote maximally-entangled pairs. 
(a) SU(4) with 20-dimensional representation and 
(b) SU(6) with 175-dimensional representation. 
\label{fig:VB-construction}}
\end{figure}

To get a good insight into the properties of the GS of this SU($N$) Heisenberg chain,
we construct a series of model GS, the VBS states~\cite{AKLT},
whose parent Hamiltonian is close to the original one. 
We start from a pair of the self-conjugate representations (characterised by a Young tableau 
with {\em one} column and $N/2$ rows) on each site and create maximally-entangled pairs 
between adjacent sites (see fig.~\ref{fig:VB-construction}).  
Last, we obtain the model VBS state by projecting the tensor-product 
states on each site onto the desired self-conjugate representation~\cite{remarkVBS}.  
We explicitly constructed the matrix-product representation 
\begin{equation}
\sum_{\{m_{i}\}}A_{1}(m_1)A_{2}(m_2)\cdots A_{i}(m_i)\cdots |m_1,m_2,\ldots,m_i
,\ldots\rangle
\end{equation}
of such a state for $N=4$ 
(with the dimension of the $A$-matrices being 6 and $\{m_i\}$ labelling the 20 physical states at each 
site) to obtain `spin-spin' correlations exponentially decaying 
with correlation length $1/\ln 5\approx 0.6213$.   
The parent Hamiltonian which supports the above VBS state as the exact GS 
is given by
\begin{eqnarray}
{\cal H}_{\rm VBS}  &=& J_s \sum_{i}\Big\{ S^{A}_{i} S^{A}_{i+1} + \frac{13}{108}(S^{A}_{i} S^{A}_{i+1})^2 
 \nonumber \\&&
 + \frac{1}{216}(S^{A}_{i} S^{A}_{i+1})^3 \Big\},
\label{VBSmodel}
\end{eqnarray}
where $S^{A}_{i}$ denote the SU($4$) spin operators in the 20-dimensional representation.
We observe that model (\ref{VBSmodel}) is not very far from the original pure Heisenberg 
Hamiltonian (with spin exchange $J_s$) obtained by the $t/U$-expansion. 
This strongly suggests that an SU(4) topological phase
is stabilised in the strong-coupling regime with the emergent edge states belonging to the 6-dimensional 
representation of SU(4).
From the structure of the edge states, one can easily see that the VBS state found above 
belongs to one of the $N$ topological classes protected by SU($N$) symmetry \cite{quella}. 

It is interesting to check its robustness in the absence of SU($N$) symmetry. 
Recently, it has been demonstrated that the even-fold degenerate structure in the entanglement 
spectrum is a fingerprint of the topological Haldane phase protected by inversion 
symmetry~\cite{pollman10}.
One may think that the $N=4$ VBS state above represents a topologically 
robust Haldane state since its six finite entanglement eigenvalues are all degenerate.  
However, this even-fold degeneracy is accidental 
and the corresponding topological phase is protected only 
in the presence of high symmetries [e.g. SU($N$)]. 
In fact, if the phase is protected by the link-inversion symmetry, 
the unitary matrix $U_{I}$ satisfying~\cite{Garcia-W-S-V-C-08}
\begin{equation}
   A^{T}(m) = e^{i\theta_{I}}{U_{I}}^{\dagger} A(m) U_{I} 
   \label{unitary_matrix}
\end{equation}
should be antisymmetric~\cite{pollman10}. 
For the $N=4$ VBS state, ${U_{I}}^{T}=U_{I}$ and 
the state becomes trivial only in the presence of such an elementary symmetry as the link-inversion. 
Since the dimension of the self-conjugate representation 
is rapidly growing (1764 for SU(8)), it is not practical to explicitly construct the MPS for $N\geq 6$.   
However, it can be shown that the symmetry of $U_{I}$ is determined solely by that  
of the maximally-entangled (singlet) pair; if the latter pair is antisymmetric with respect to 
the interchange of the two constituent states, $U_{I}$ is antisymmetric and the corresponding MPS remains 
topological even without high symmetries.  

By investigating the form of the maximally-entangled pair, we conclude that for $N/2$ odd 
the VBS state remains to be the stable topological Haldane phase 
protected by inversion symmetry while it is not the case for $N/2$ even. 
In particular, as in the HO case, 
we expect that the gapped SU($N$) phase realised along the line 
$J=2NU/(N+2)$ is SPT when $I=1/2,5/2,9/2, \ldots$.  

\section{Low-energy approach}  
The low-energy effective field theory of the lattice model
(\ref{alkaTmodel}) is derived by expressing the standard continuum limit of 
the lattice fermionic operators $c_{l\alpha\,i}$ in terms of $2N$
left- and right-moving $L_{l\alpha},\,R_{l\alpha}$ Dirac fermions \cite{bookboso}: 
$c_{l\alpha\,i} \sim R_{l \alpha} e^{ik_F x}+ L_{l \alpha} e^{-i k_F x}$, 
where $x=ia_0$, $a_0$ being the lattice spacing and $k_F = \pi/(2 a_0)$ the Fermi momentum. 
The continuum Hamiltonian then takes the form of $2N$ Dirac fermions 
coupled with marginal four-fermions interactions. 
The low-energy properties of the resulting field theory
are determined by means of a one-loop renormalization group (RG) analysis.
This analysis has been done in the $N=2$ case~\cite{nonneViet,nonne2010}.
For instance, four different fully gapped Mott-insulating phases have been found when $J=J_z$.
On top of conventional two-fold degenerate phases with spin-Peierls (SP) and charge-density wave (CDW) orderings, 
the pseudospin-1 HO and the spin-1 Haldane phases, which have been identified in the strong-coupling
analysis, persist in the weak-coupling regime as well. 

The RG analysis in the general $N>2$ case is much more involved.
In some regions of the phase diagram, we find that the one-loop RG flow is attracted 
along two isotropic rays with SO($4N$) symmetry which is the maximal continuous symmetry achievable 
for $2N$ Dirac fermions.
These highly-symmetric rays signal the emergence of the CDW and SP phases
in the general $N$ case~\cite{nonne2011}. 
In sharp contrast to $N=2$, the one-loop RG flow for $N>2$ features a region
with no symmetry restoration in the infrared (IR) limit. 
There is a separation of the energy scales: one of the perturbations, which depends only
on the SU($N$) spin degrees of freedom, 
reaches the strong-coupling regime faster than the others.
A spin gap $\Delta_s$ is thus formed among the nuclear spin degrees of freedom.
Below the energy scale of the spin gap $E \ll \Delta_s$, the dominant part of the effective
interacting Hamiltonian only involves the remaining charge and orbital
degrees of freedom and, when $J=J_z$, it reads as follows:
 \begin{eqnarray}
  \mathcal{H}^{\rm eff}_{\rm int}  \simeq
    \lambda   \left(  {\rm Tr} g \right)^2 
       + \mu \cos \left(\sqrt{8 \pi K_c /N} \Phi_c \right) ,
  \label{effhamregion3}
\end{eqnarray}
where $\Phi_c$ is a Bose field which accounts for the  U(1)$_c$
charge degrees of freedom, and $K_c$ is its Luttinger parameter~\cite{bookboso}. 
The low-energy properties of the charge degrees of freedom are captured
by the sine-Gordon model at $\beta^2 = 8 \pi K_c /N$ and we expect that the charge sector 
is gapped away since  $K_c <N$. The IR properties of model (\ref{effhamregion3}) thus
depend only on the orbital degrees of freedom and are 
described by the SU(2)$_N$ conformal field theory (CFT) perturbed by its spin-1
operator $ \left(  {\rm Tr} g \right)^2$ ($g$ being the SU(2) matrix) with scaling dimension $4/(N+2)$.
In this respect, the resulting low-energy field theory is exactly that of the spin-$N/2$ SU(2)
Heisenberg chain derived by Affleck and Haldane using the non-Abelian bosonization
approach~\cite{affleckhaldane}.
We thus deduce the emergence of a Haldane-gap phase when $N$ is even,
i.e. for general half-integer nuclear spins.  
The resulting Haldane phase is identified with the HO phase which have been found 
already by the strong-coupling approach and is a collective singlet state formed among 
the orbital degrees of freedom.

In stark contrast, within the weak-coupling approach, we could see no evidence for 
a similar SU($N$) topological phase of the nuclear spins for $N>2$.  
In particular, along the $J = 2 NU/(N+2) > 0$  line ($N$ being even), a SP phase is found in the 
RG approach instead of  an SU($N$) topological phase expected from the strong-coupling
argument. 
Therefore a quantum phase transition necessarily occurs at intermediate couplings 
and it is tempting to conjecture that the quantum critical
point is described by an SU($N$)$_2$ CFT.  
The situation may be understood as follows. 
First, one can determine by symmetry the low-energy field theory which is valid in the vicinity of this 
SU($N$)$_2$ quantum critical point:
$\mathcal{H}_{\rm eff} \simeq \mathcal{H}_{SU({\rm N})_2} + (g -g_c) |{\rm Tr} \; G|^2$,
$G$ being the SU($N$)$_2$ primary field which belongs to the fundamental representation
of  SU($N$).
Following Ref.~\cite{afflecksun}, the nature of the phases of the latter model
can be inferred from a semiclassical approach; 
when $g<g_c$, one has $ \langle {\rm Tr} G \rangle \ne 0$, 
and the GS is two-fold degenerate as a consequence of 
broken translation symmetry ($G \rightarrow - G$).  
This may be identified with the SP phase found in the weak-coupling limit.
On the strong coupling side, when $g > g_c$, 
the semiclassical analysis now gives an SU($N$) matrix with
the constraint $ {\rm Tr}  G =0$, and the phase is translationally invariant. 
The resulting effective field theory is known to be 
the Grassmannian sigma model on $U(N)/[U(N/2)\times U(N/2)]$ manifold
with a $\theta = 2 \pi$, topological theta term, 
which is massive~\cite{afflecksun}. The latter is known to be the semiclassical field theory
of the SU($N$) Heisenberg spin chain in self-conjugate representations with two columns~\cite{sachdev}.
We thus conclude that the SU($N$)  topological phase, identified within the 
VBS approach based on the strong-coupling Hamiltonian, emerges for $g > g_c$. 
 
\begin{figure}
\onefigure[width=\linewidth,clip]{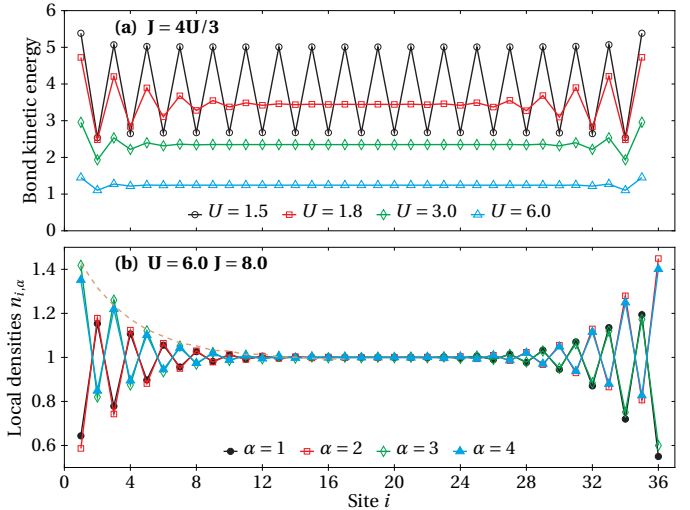}
\caption{(Colour on-line) (a) Bond kinetic energy for various $U>0$ along the line $J=4U/3$ ($N=4$) in a $L=36$ chain
($t=1$).
(b) local densities $n_{i,\alpha}$  in the topological phase
showing the existence of edge states. The dashed line is an exponential fit with a correlation length of 
around 3 lattice spacings. 
\label{fig:figDMRG.N=4}}
\end{figure}

\section{DMRG calculations}  A density-matrix renormalization group~\cite{DMRG} (DMRG) 
is clearly called for to shed light on the existence of this quantum phase transition 
for moderate couplings. 
Model (\ref{alkaTmodel}) was rewritten as an $N$-leg Hubbard ladder
with additional rung interactions to get more efficient
simulations. 
Typically, we used open boundary conditions and kept up to 1600 and 3200 states for
convergence with $N=2$ and $N=4$ respectively to get a discarded weight below $10^{-5}$.

Fig.~\ref{fig:figDMRG.N=4}(a) shows the bond kinetic energy for
various $U>0$ using $N=4$ along the special line $J= 2 NU/(N+2)=4U/3$.
As $U$ increases, we clearly see a quantum phase
transition from a SP phase to a uniform non-degenerate one. 
This is in agreement with our conjecture based on the low-energy and
strong-coupling analysis. 
In order to confirm that the non-degenerate phase
corresponds to the SU(4) topological phase discussed above, we plot in
fig.~\ref{fig:figDMRG.N=4}(b) the local densities
$n_{i,\alpha}=\sum_{l} c_{l \alpha,\,i}^\dag c_{l \alpha,\,i}^{}$ for
each flavour $\alpha=1,\ldots,4$ in this phase to investigate its edge states. 
Clearly, edge states
are present with an exponential profile and a correlation length of
the order of 3 lattice spacings for $U=6$ and $J=8$ (see
fit). Moreover, DMRG data randomly show different kinds of edge
states: in the sense that for a given edge, two flavours out of four
give the same local densities~\cite{comment}. 
Since there are six ways to choose two flavours out of four, our
numerical data confirm that edge states are 6-fold degenerate in
agreement with the VBS prediction that they belong to 6-dimensional SU(4)
representation. 

Another related evidence for the SU($N$) topological phase comes from the degeneracy of the entanglement spectrum.  
According to the classification of the SU($N$) SPT phases in~\cite{quella}, the SU(4) VBS state 
discussed above corresponds to one of the stable topological phases predicted there.  
Therefore, one may expect, on physical grounds, that the six-fold degeneracy of 
the lowest entanglement eigenvalue can be used as the fingerprint. 
To check this, we considered the ground-state corresponding to the parameters used in fig.~\ref{fig:figDMRG.N=4}(b) 
and measured the entanglement spectrum in the SU(4) topological phase. 
In fact, we observed that the lowest eigenvalue (i.e. the largest weight) is 6-fold degenerate as expected.  
The full characterisation of our SU($N$) topological phase in terms of entanglement spectrum 
and the (non-local) order parameters~\cite{non-local-OP1,non-local-OP2,non-local-OP3} 
will be discussed in future work.

We also observed that these edge states are absent in the SP phase, 
and correspondingly the entanglement spectrum is non-degenerate. Therefore, this quantum phase transition is an example of topological one since the entanglement spectrum is totally different in each phase. However, the precise location of the phase transition and its universality class, as well as the full phase diagram, require 
extensive large-scale simulations and are thus beyond the scope of the Letter.
Finally, let us mention that our preliminary data indicate that this topological phase has a rather large extension in the phase diagram (a large part of the quadrant $U,J>0$), which confirms its robustness and relevance for such parameters. 

\begin{figure}
\onefigure[width=\linewidth,clip]{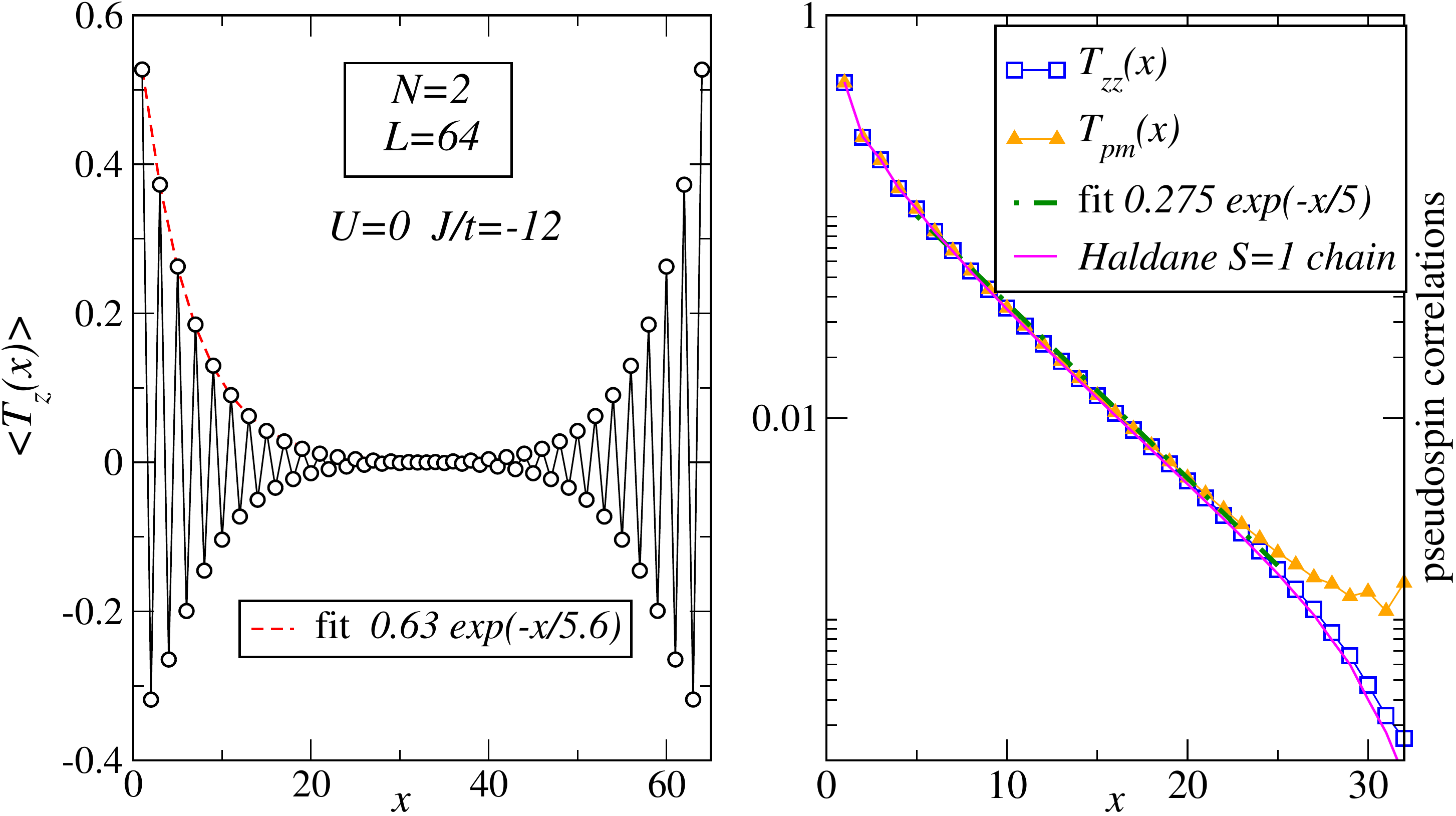}
\caption{(Colour on-line) $N=2$ data for a chain of length $L=64$ with $U=0$ and $J/t=-12$ 
(a) local average $\langle T_z(x)\rangle$ vs position $x$ in the GS 
with total $T_z=1$ showing evidence of localised edge states; 
(b) pseudospin correlations $T_{zz}(x)=\langle T_z(L/2) T_z(L/2+x)\rangle$ and $T_{pm}(x)=\langle T_+(L/2) T_-(L/2+x)\rangle/2$ 
in the GS with $T_z=0$ exhibit an $SU(2)_o$ symmetry, are short-ranged and almost identical to spin correlations measured in spin-1 Heisenberg chain. 
Bulk and edge correlation lengths are close to 5 lattice spacings. 
\label{fig:figDMRG.HO}}
\end{figure}

Now, we turn to the illustration of the HO phase for
$N=2$ since the case $N=4$ would be quite demanding numerically~\cite{spin2}. Fig.~\ref{fig:figDMRG.HO}(a) 
shows the presence of edge states in a finite chain of length $L=64$, similarly to what is found in 
the Haldane phase of the spin-1 chain. In fig.~\ref{fig:figDMRG.HO}(b), we also plot 
pseudospin correlations (taken from the middle of the chain) that are short-ranged and almost identical to spin correlations measured in a spin-1 Heisenberg chain. 
Overall, we confirm 
the existence of the HO phase in this region.

\section{Conclusion} 

We have found that nontrivial topological phases can exist in 1D alkaline-earth cold atomic systems, 
thanks to the interplay between orbital and nuclear-spin degrees of freedom.
More explicitly, we have shown the existence of (i) the \emph{Haldane-orbital phase}, which is 
the analogue of the usual Haldane phase but with orbital degrees of freedom, and also (ii) an 
SU($N$) topological phase for which we give an explicit construction using MPS approach and that 
possesses nontrivial edge states. 
Moreover, both topological phases correspond to the symmetry-protected Haldane phase 
provided $N/2$ is odd, 
which is directly relevant to strontium and ytterbium cold atoms, while the new SU($N$) phase 
for the other even-$N$s is stable only in the presence of exact SU($N$) symmetry. 

In principle, experimental realisations of such phases could be achieved using alkaline-earth cold atoms such 
as ytterbium (${}^{171}$Yb and ${}^{173}$Yb)~\cite{taie,taie_2012} 
and strontium (${}^{87}$Sr)~\cite{desalvo}, 
that are known to realise SU($N$) symmetry with great accuracy. 
In such a case, the orbital degree of freedom could be 
provided either by the metastable excited $e$ state 
of the atoms \cite{Gorshkov2010}, or by the $p$-band of the 1D optical lattice~\cite{kobayashi}.
In a harmonic trap potential, the polarisation on the outer region of cold atomic systems 
can be measured \cite{Partridge_2006, Liao_2010}  which may 
open perspectives for the observation of SPT in alkaline-earth cold atoms.

\acknowledgments
The authors would like to thank E. Boulat and T. Koffel for useful discussions. 
Numerical simulations were performed at CALMIP. 
Supported in part at the Technion by a fellowship from the Lady Davis Foundation.

\bibliographystyle{eplbib}

\end{document}